\documentclass[sigconf]{acmart}

\usepackage{algorithm}
\usepackage{multirow}
\usepackage{enumitem}
\usepackage{bm}
\usepackage{xspace}
\usepackage{marvosym}

\usepackage{bbding}
\usepackage{subcaption}
\usepackage{tabularx}
\AtBeginDocument{%
  }

\begin{document}

\title[Sparse Meets Dense: Unified Generative Recommendations with Cascaded Sparse-Dense Representations]{\texorpdfstring{Sparse Meets Dense: \\Unified Generative Recommendations with Cascaded Sparse-Dense Representations}{Sparse Meets Dense: Unified Generative Recommendations with Cascaded Sparse-Dense Representations}}

\author{Yuhao Yang}
\email{yangyuhao01@baidu.com}
\affiliation{
    \institution{Baidu Inc.}
    \city{Beijing}
    \country{China}
}

\author{Zhi Ji}
\email{jizhi@baidu.com }
\affiliation{
    \institution{Baidu Inc.}
    \city{Beijing}
    \country{China}
}

\author{Zhaopeng Li}
\email{lizhaopeng@baidu.com}
\affiliation{
    \institution{Baidu Inc.}
    \city{Beijing}
    \country{China}
}

\author{Yi Li}
\email{liyi01@baidu.com}
\affiliation{
    \institution{Baidu Inc.}
    \city{Beijing}
    \country{China}
}

\author{Zhonglin Mo}
\email{mozhonglin@baidu.com }
\affiliation{
    \institution{Baidu Inc.}
    \city{Beijing}
    \country{China}
}

\author{Yue Ding}
\email{dingyue03@baidu.com}
\affiliation{
    \institution{Baidu Inc.}
    \city{Beijing}
    \country{China}
}

\author{Kai Chen}
\email{chenkai23@baidu.com}
\affiliation{
    \institution{Baidu Inc.}
    \city{Beijing}
    \country{China}
}

\author{Zijian Zhang}
\email{zhangzijian02@baidu.com}
\affiliation{
    \institution{Baidu Inc.}
    \city{Beijing}
    \country{China}
}

\author{Jie Li}
\email{lijie06@baidu.com}
\affiliation{
    \institution{Baidu Inc.}
    \city{Beijing}
    \country{China}
}

\author{Shuanglong Li}
\email{lishuanglong@baidu.com}
\affiliation{
    \institution{Baidu Inc.}
    \city{Beijing}
    \country{China}
}

\author{Lin Liu}
\email{liulin03@baidu.com}
\affiliation{
    \institution{Baidu Inc.}
    \city{Beijing}
    \country{China}
}

\renewcommand{\authors}{Yuhao Yang, Zhi Ji, Shuanglong Li, Lin Liu}
\renewcommand{\shortauthors}{Yang, et al.}

\newcommand{\ie}{\emph{i.e.,}\xspace}
\newcommand{\eg}{\emph{e.g.,}\xspace}
\newcommand{\aka}{\emph{a.k.a.,}\xspace}
\newcommand{\etal}{\emph{et al.}\xspace}
\newcommand{\paratitle}[1]{\vspace{1.5ex}\noindent\textbf{#1}}
\newcommand{\wrt}{w.r.t.\xspace}
\newcommand{\ignore}[1]{}

\newcommand{\tba}{\textcolor{red}{xxx }}
\newcommand{\outd}{\textcolor{red}{[Outdated]}~}
\newcommand{\tabincell}[2]{\begin{tabular}{@{}#1@{}}#2\end{tabular}}

\definecolor{dark2green}{rgb}{0.1, 0.65, 0.3}
\definecolor{dark2orange}{rgb}{0.9, 0.4, 0.}
\definecolor{dark2purple}{rgb}{0.4, 0.4, 0.8}
\newcommand{\first}[1]{\textbf{#1}}
\newcommand{\second}[1]{\underline{#1}}
\newcommand{\third}[1]{\textbf{\textcolor{dark2purple}{#1}}}

\begin{abstract}
Generative models have recently gained attention in recommendation systems by directly predicting item identifiers from user interaction sequences. However, existing methods suffer from significant information loss due to the separation of stages such as quantization and sequence modeling, hindering their ability to achieve the modeling precision and accuracy of sequential dense retrieval techniques. Integrating generative and dense retrieval methods remains a critical challenge. To address this, we introduce the Cascaded Organized Bi-Represented generAtive retrieval (COBRA) framework, which innovatively integrates sparse semantic IDs and dense vectors through a cascading process. Our method alternates between generating these representations by first generating sparse IDs, which serve as conditions to aid in the generation of dense vectors. End-to-end training enables dynamic refinement of dense representations, capturing both semantic insights and collaborative signals from user-item interactions. During inference, COBRA employs a coarse-to-fine strategy, starting with sparse ID generation and refining them into dense vectors via the generative model. We further propose BeamFusion, an innovative approach combining beam search with nearest neighbor scores to enhance inference flexibility and recommendation diversity. Extensive experiments on public datasets and offline tests validate our method's robustness. Online A/B tests on a real-world advertising platform with over 200 million daily users demonstrate substantial improvements in key metrics, highlighting COBRA's practical advantages.
\end{abstract}


\maketitle

\section{Introduction}

\begin{figure*}[htbp]
\centering
\includegraphics[width=1.00\textwidth]{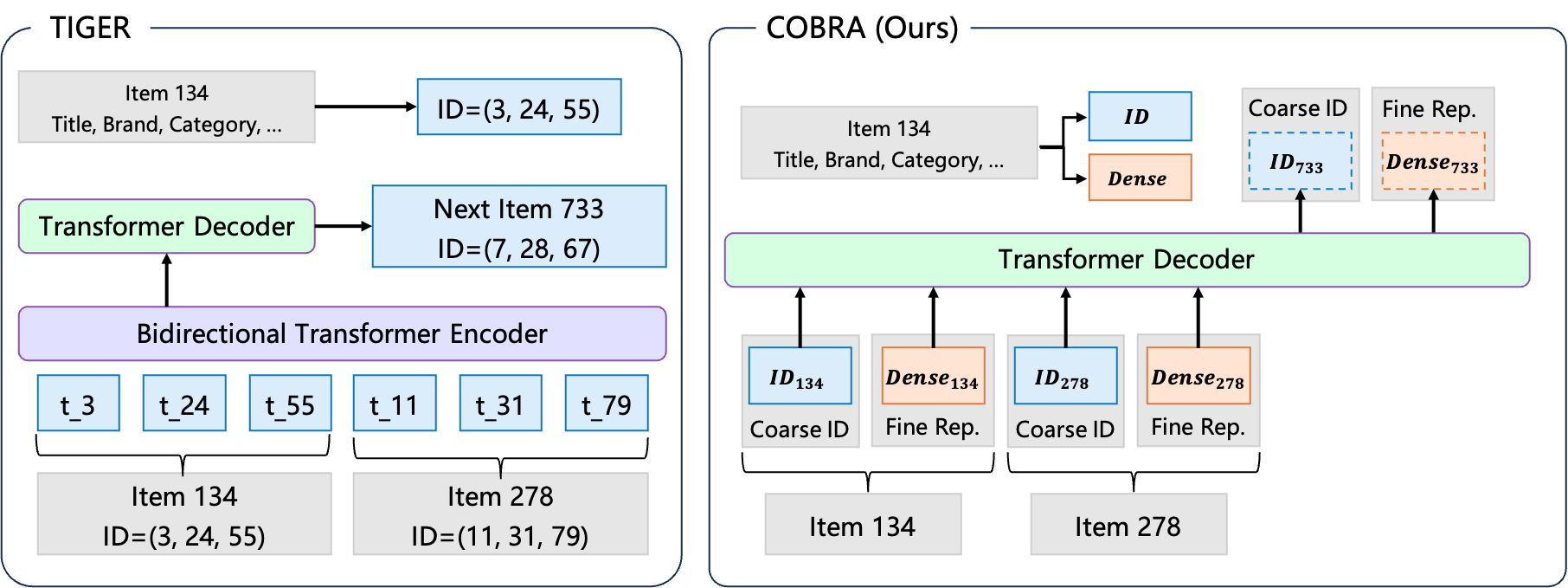}
\caption{Comparison of generative recommendation paradigms. The left section illustrates traditional generative retrieval approaches, exemplified by TIGER, which utilize a sequence of sparse IDs as input within a Transformer encoder-decoder architecture to directly predict the sparse ID of the next item. The right section depicts the proposed COBRA framework, which employs Cascaded Organized Bi-Represented generAtive retrieval. This approach integrates sparse IDs to capture coarse-grained semantic information and dense vectors to encapsulate fine-grained detail. The cascaded representation is processed by a Transformer decoder that sequentially predicts the sparse ID followed by the dense vector.}
\label{fig:paradigm_comparison}
\end{figure*}
Recommendation systems are vital components of modern digital ecosystems, providing personalized item suggestions that align with user preferences across e-commerce platforms, streaming services, and social networks~\cite{deepfm,wide_and_deep,bridging_search_and_recommendation}. Recent advancements have focused on sequential recommendation methods, which leverage the sequential nature of user interactions to enhance recommendation performance~\cite{autoint, bst,scaling_law_large_sequential_recommendation,tim4rec}. Notable models like SASRec~\cite{Sasrec} and BERT4Rec~\cite{bert4rec} have demonstrated the effectiveness of sequence models in capturing user behavior patterns.

The emergence of generative models has further expanded the capabilities of recommendation systems~\cite{hstu,openp5,hllm}. Unlike traditional sequential recommendation methods, generative models can directly predict target items based on user behavior sequences~\cite{gen_recsys_review,cost,better_generalization_with_case_study}. These models handle complex user-item interactions and offer emerging abilities such as reasoning and few-shot learning, which significantly improve recommendation accuracy and diversity~\cite{dlcrec,larr,secor}. Among these, TIGER~\cite{tiger} is a pioneering approach in generative retrieval for recommendation systems. As depicted in Figure~\ref{fig:paradigm_comparison}(Lower Left), TIGER leverages a Residual Quantized Variational AutoEncoder (RQ-VAE)~\cite{rq_vae} to encode item content features into hierarchical semantic IDs, allowing the model to share knowledge across semantically similar items without the need for individual item embeddings. Beyond TIGER, several other methods have been proposed to further explore the integration of generative models with recommendation systems. LC-Rec~\cite{lc-rec} aligns semantic and collaborative information using RQ-VAE with a series of alignment tasks. ColaRec~\cite{colarec} combines collaborative filtering signals with content information by deriving generative identifiers from a pretrained recommendation model. IDGenRec~\cite{idgenrec} leverages large language models to generate unique, concise, and semantically rich textual identifiers for recommended items, showing strong potential in zero-shot settings. 

Despite these innovations, existing generative recommendation methods still face several challenges compared to sequential dense retrieval methods~\cite{survey_gen_search_and_recommendation,recommendation_with_gen_model}. Sequential dense retrieval methods, which rely on dense embeddings for each item, offer high accuracy and robustness but require substantial storage and computational resources. In contrast, generative methods, while efficient, often struggle with fine-grained similarity modeling~\cite{liger}. To effectively leverage the strengths of both retrieval paradigms, we propose Cascaded Organized Bi-Represented generAtive retrieval(COBRA), a framework that synergizes generative and dense retrieval. Figure~\ref{fig:paradigm_comparison}(Right) illustrates the cascaded sparse-dense representations in COBRA. The proposed method introduces a cascaded generative retrieval framework alternating between generating sparse IDs and dense vectors. This approach mitigates information loss inherent in ID-based methods. Specifically, COBRA's input is a sequence of cascaded representations composed of sparse IDs and dense vectors corresponding to items in the user's interaction history. During training, the dense representations are learned through contrastive learning objectives in and end-to-end manner. By first generating the sparse ID and then the dense representation, COBRA reduces the learning difficulty of dense representations and promotes mutual learning between the two representations. During inference, COBRA employs a coarse-to-fine generation process, starting with sparse ID that provides a high-level categorical sketch capturing the categorical essence of the item. The generated ID is then appended to the input sequence and fed back into the model to predict the dense vector that captures the fine-grained details, enabling more precise and personalized recommendations. To ensure flexible inference, we introduce BeamFusion, a sampling technique combining beam search with nearest neighbor retrieval scores, ensuring controllable diversity in the retrieved items. Unlike TIGER, which relies solely on sparse IDs, COBRA harnesses the strengths of both sparse and dense representations.

Our main contributions are as follows:
\begin{itemize}
\item \textbf{Cascaded Bi-Represented Retrieval Framework}: We introduce COBRA, a novel cascading framework that alternates between generating sparse semantic IDs and dense vectors. By incorporating dense representations into the ID sequence, COBRA supplements the information loss inherent in ID-based methods. Using sparse IDs as conditions for generating dense vectors reduces the learning difficulty of dense representations.
\item \textbf{Learnable Dense Representations via End-to-End Training}: COBRA leverages the original item data as input to generate dense representations through end-to-end training. Unlike static embeddings, COBRA's dense vectors are dynamically learned, capturing semantic information and fine-grained details.
\item \textbf{Coarse-to-Fine Generation Process}: During inference, COBRA employs a coarse-to-fine generation process. It first generates sparse IDs, which are then fed back into the model to produce refined dense representations, enhancing the granularity of the dense vectors. Additionally, we propose BeamFusion for flexible and diverse recommendation.
\item \textbf{Comprehensive Empirical Validation}: Through extensive experiments on multiple benchmark datasets, we demonstrate that COBRA achieves superior performance in terms of recommendation accuracy compared to existing state-of-the-art methods. These results validate the effectiveness of COBRA and its ability to balance precision and diversity in recommendation tasks.
\end{itemize}
\section{Related Work}
\textbf{Sequential Dense Recommendation.} Sequential dense recommendation systems leverage user interaction sequences to learn dense representations for users and items~\cite{Sasrec,s3rec,recformer}, capturing both long-term preferences and short-term dynamics~\cite{mojito,duorec,stosa,automlp}. Early models such as GRU4Rec~\cite{gru4rec} utilized Recurrent Neural Networks (RNNs)~\cite{rnn} to capture temporal dependencies in user behavior. Caser~\cite{caser} applied Convolutional Neural Networks (CNNs)~\cite{cnn} to treat sequences as "images" for spatial feature extraction. The advent of Transformer-related models, such as SASRec~\cite{Sasrec} and BERT4Rec~\cite{bert4rec}, has significantly advanced this field. These models employ self-attention mechanisms to capture complex user behaviors, with SASRec focusing on autoregressive tasks and BERT4Rec on bidirectional context modeling. More advanced models like PinnerFormer~\cite{pinnerformer} and FDSA~\cite{fdsa} have further enhanced user representation by leveraging Transformers for long-term behavior modeling and feature integration. Recent works, including ZESRec~\cite{zesrec}, UniSRec~\cite{unisrec}, and RecFormer~\cite{recformer}, have emphasized cross-domain transferability by incorporating textual features and employing contrastive learning techniques. RecFormer, in particular, has unified language understanding and sequence recommendation through bidirectional Transformers.

\textbf{Generative Recommendation.} The increasing popularity of generative models across various domains has led to a paradigm shift in recommendation systems from discriminative to generative models~\cite{genrec,mbgen,calrec,eager,bigrec,transrec}. Generative models directly generate item identifiers rather than computing ranking scores for each item~\cite{tiger,how2index,letter,mmgrec}. P5~\cite{p5} has transformed various recommendation tasks into natural language sequences, providing a universal framework for recommendation completion through unique training objectives and prompts. TIGER~\cite{tiger} pioneered the application of generative retrieval to recommendations by using a residual quantized autoencoder to create semantically rich indexing identifiers. These identifiers are then utilized by a transformer-based model to generate item identifiers from user histories. LC-Rec~\cite{lc-rec} enhanced this approach by aligning semantic identifiers with collaborative filtering techniques through additional alignment tasks. IDGenRec~\cite{idgenrec} merged generative systems with large language models to generate unique, semantically dense textual identifiers, demonstrating strong performance even in zero-shot settings. SEATER~\cite{seater} focused on maintaining semantic consistency through balanced k-ary tree-structured indexes refined by contrastive and multi-task learning. ColaRec~\cite{colarec} aligned content-based semantic spaces with collaborative interaction spaces to improve recommendation efficacy.
However, existing generative methods face several challenges. For instance, methods based on discrete IDs may lack fine-grained details and suffer from information loss, which can limit their ability to accurately capture user preferences~\cite{flip}. Additionally, approaches that rely on natural language may struggle to align linguistic expressions with the requirements of recommendation tasks, potentially leading to suboptimal performance~\cite{llm_srec}. To address these issues, LIGER~\cite{liger} proposes a hybrid model that combines advantages of generative and dense retrieval methods. It simultaneously generates sparse IDs and dense representations, treating them as complementary representations of the same object granularity. This hybrid approach effectively narrows the gap between generative and dense retrieval methods to some extent. However, LIGER's IDs and dense representations share the same granularity, and the dense representations are pre-trained and fixed. Therefore, how to more flexibly combine generative and dense retrieval methods is still an open question that needs further exploration.

\section{Methodology}
\begin{figure*}[ht]
    \centering
    \includegraphics[width=1.00\textwidth]{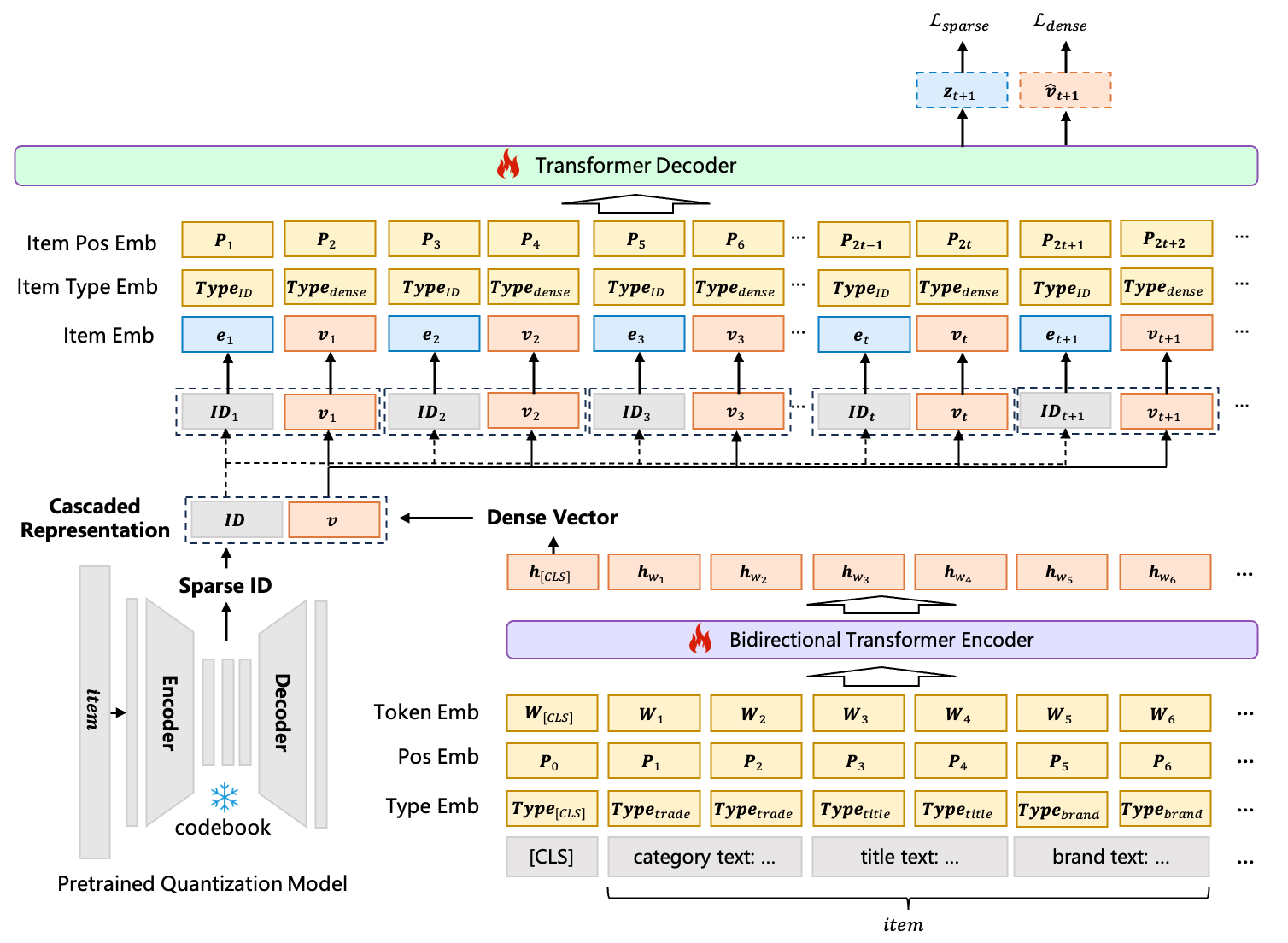}
    \caption{The architecture of COBRA. The model employs a cascaded sparse-dense representation approach, where sparse IDs are generated via Residual Quantization and dense vectors are produced by a trainable Transformer Encoder. These representations serve as inputs to a Transformer Decoder, which alternates between predicting sparse IDs and dense vectors. The predicted outputs are used to compute the loss functions $\mathcal{L}_{\text{sparse}}$ and $\mathcal{L}_{\text{dense}}$. For the sake of simplicity, the figure illustrates an example with a single level of sparse ID.}
    \label{fig:overall}
\end{figure*}

This section introduces the Cascaded Organized Bi-Represented generAtive Retrieval (COBRA) framework, which integrates cascaded sparse-dense representations and coarse-to-fine generation to enhance recommendation performance. Figure~\ref{fig:overall} illustrates the overall framework of COBRA.

\subsection{Sparse-Dense Representation}

\subsubsection{Sparse Representation}
COBRA generates sparse IDs using a Residual Quantized Variational Autoencoder (RQ-VAE), inspired by the approach in TIGER~\cite{tiger}. For each item, we extract its attributes to generate a textual description, which is embedded into a dense vector space and quantized to produce sparse IDs. These IDs capture the categorical essence of items, forming the basis for subsequent processing. For the sake of brevity, the subsequent methodology descriptions will assume that the sparse ID consists of a single level. However, it should be noted that this approach can be easily extended to accommodate scenarios involving multiple levels.

\subsubsection{Dense Representation}
To capture nuanced attribute information, we develop an end-to-end trainable dense encoder, encoding item textual contents. Each item's attributes are flattened into a text sentence, prefixed with a [CLS] token, and fed into a Transformer-based text encoder $\textbf{Encoder}$. The dense representation $\mathbf{v}_t$ is extracted from the output corresponding to the [CLS] token, capturing fine-grained details of the item's textual content. As illustrated in the lower part of Figure~\ref{fig:overall}, we incorporate position embeddings and type embeddings to model the positional and context of tokens within the sequence. These embeddings are added to the token embeddings in an additive manner, enhancing the model's ability to distinguish between different tokens and their positions in the sequence.

\subsubsection{Cascaded Representation}
The cascaded representation integrates sparse IDs and dense vectors within a unified generative model. Specifically, for each item, we combine its sparse ID  $ID_t$ and dense vector $\mathbf{v}_t$ to form a cascaded representation $(ID_t, \mathbf{v}_t)$. This approach leverages the strengths of both representations, providing a more comprehensive characterization of items: sparse IDs provide a stable categorical foundation through discrete constraints, while dense vectors maintain continuous feature resolution, ensuring that the model captures both high-level semantics and fine-grained details.

\subsection{Sequential Modeling}

\subsubsection{Probabilistic Decomposition}
The probability distribution modeling of the target item is factorized into two stages, leveraging the complementary strengths of sparse and dense representations. Specifically, instead of directly predicting the next item $s_{t+1}$ based on the historical interaction sequence $S_{1:t}$, COBRA predicts the sparse ID $ID_{t+1}$ and the dense vector $\mathbf{v}_{T+1}$ separately:
\begin{equation}
P(ID_{t+1}, \mathbf{v}_{t+1} | S_{1:t}) = P(ID_{t+1} | S_{1:t}) P(\mathbf{v}_{t+1} | ID_{t+1}, S_{1:t})
\end{equation}
where $P(ID_{t+1} | S_{1:t})$ represents the probability of generating the sparse ID $ID_{t+1}$ based on the historical sequence $S_{1:t}$, capturing the categorical essence of the next item. $P(\mathbf{v}_{t+1} | ID_{t+1}, S_{1:t})$ represents the probability of generating the dense vector $\mathbf{v}_{t+1}$ given the sparse ID $ID_{t+1}$ and the historical sequence $S_{1:t}$, capturing the fine-grained details of the next item. This decomposition allows COBRA to leverage both the categorical information provided by sparse IDs and the fine-grained details captured by dense vectors.

\subsubsection{Sequential Modeling with a Unified Generative Model}

For sequential modeling, we utilize a unified generative model based on the Transformer architecture to effectively capture sequential dependencies in user-item interactions. The Transformer receives an input sequence of cascaded representations, with each item represented by its sparse ID and dense vector.

\textbf{Embedding Sparse IDs}  
The sparse ID, denoted as $ID_t$, is transformed into a dense vector space through an embedding layer: $\mathbf{e}_{t} = \textbf{Embed}(ID_t)$. This embedding $\mathbf{e}_{t}$ is concatenated with the dense vector $\mathbf{v}_t$ to form the model's input at each time step:
\begin{equation}
\mathbf{h}_t = [\mathbf{e}_{t}; \mathbf{v}_t]
\end{equation}

\textbf{Transformer Modeling}  
Our Transformer Decoder model comprises multiple layers, each featuring self-attention mechanisms and feedforward networks. As depicted in the upper part of Figure~\ref{fig:overall}, the input sequence to the Decoder consists of cascaded representations. To enhance modeling of sequential and contextual information, these representations are augmented with item position and type embeddings. For brevity, mathematical formulations in the following sections focus on the cascaded sequence representation, omitting explicit notation for position and type embeddings. The Decoder processes this enriched input to generate contextualized representations for predicting the subsequent sparse ID and dense vector.

\textbf{Sparse ID Prediction}  
Given history interaction sequence $S_{1:t}$, to predict the sparse ID $ID_{t+1}$, the Transformer input sequence is:
\begin{equation}
\begin{aligned}
\mathbf{S}_{1:t} &= [\mathbf{h}_1, \mathbf{h}_2, \ldots, \mathbf{h}_t] \\
&= [\mathbf{e}_{1}, \mathbf{v}_1, \mathbf{e}_{2}, \mathbf{v}_2, \ldots, \mathbf{e}_{t}, \mathbf{v}_{t}]
\end{aligned}
\end{equation}
where each $\mathbf{h}_i$ is a concatenation of the sparse ID embedding and the dense vector for the $i$-th item. The Transformer processes this sequence to generate contextualized representations, subsequently used to predict the next sparse ID and dense vector. Specifically, the Transformer decoder processes the sequence $\mathbf{S}_{1:t}$, producing a sequence of vectors $\mathbf{y}_t = \textbf{TransformerDecoder}(\mathbf{S}_{1:t})$. The logits for sparse ID prediction are derived as:
\begin{equation}
\mathbf{z}_{t+1} = \textbf{SparseHead}(\mathbf{y}_t)
\end{equation}
where $\mathbf{z}_{t+1}$ represents the logits for the predicted sparse ID $ID_{t+1}$.

\textbf{Dense Vector Prediction}  
For predicting the dense vector $\mathbf{v}_{t+1}$, the Transformer input sequence is: 
\begin{equation}
\begin{aligned}
\bar{\mathbf{S}}_{1:t} &= [\mathbf{S}_{1:t}, \mathbf{e}_{t+1}] \\
&= [\mathbf{e}_{1}, \mathbf{v}_1, \mathbf{e}_{2}, \mathbf{v}_2, \ldots, \mathbf{e}_{t}, \mathbf{v}_{t}, \mathbf{e}_{t+1}]
\end{aligned}
\end{equation}
The Transformer decoder processes $\bar{\mathbf{S}}_{1:t}$ to output the predicted dense vector:
\begin{equation}
\hat{\mathbf{v}}_{t+1} = \textbf{TransformerDecoder}(\bar{\mathbf{S}}_{1:t})    
\end{equation}

\subsection{End-to-End Training}

In COBRA, the end-to-end training process is designed to optimize both sparse and dense representation prediction jointly. The training process is governed by a composite loss function that combines losses for sparse ID prediction and dense vector prediction. 

The sparse ID prediction loss, denoted as $\mathcal{L}_{\text{sparse}}$, ensures the model's proficiency in predicting the next sparse ID based on the historical sequence $S_{1:t}$:
\begin{equation}
\begin{aligned}
\mathcal{L}_{\text{sparse}} &= - \sum_{t=1}^{T-1} \log\left(\frac{\exp(z_{t+1}^{ID_{t+1}})}{\sum_{j=1}^{C} \exp(z_{t+1}^j)}\right)
\end{aligned}
\end{equation}
where $T$ is the length of the historical sequence, $ID_{t+1}$ is the sparse ID corresponding to interacted item at time step $t+1$, $z_{t+1}^{ID_{t+1}}$ represents the predicted logit of groundtruth sparse ID $ID_{t+1}$ at time step $t+1$, generated by the Transformer Decoder, and $C$ denotes set of all sparse IDs.

The dense vector prediction loss $\mathcal{L}_{\text{dense}}$ focuses on refining the dense vectors, enabling them to discern between similar and dissimilar items. The loss is defined as:
\begin{equation}
\mathcal{L}_{\text{dense}} = -\sum_{t=1}^{T-1} \log \frac{\exp(\text{cos}(\hat{\mathbf{v}}_{t+1} \cdot \mathbf{v}_{t+1}))}{\sum_{{item}_{j} \in \text{Batch}} \exp(\text{cos}(\hat{\mathbf{v}}_{t+1}, \mathbf{v}_{{item}_{j}}))}
\end{equation}
where $\hat{\mathbf{v}}_t$ is the predicted dense vector, $\mathbf{v}_t$ is the ground truth dense vector for the positive item, and $\mathbf{v}_j$ represents the dense vectors of items within the batch. The term $\text{cos}(\hat{\mathbf{v}}_{t+1} \cdot \mathbf{v}_{t+1})$ represents the cosine similarity between the predicted and ground truth dense vectors. A higher cosine similarity indicates that the vectors are more similar in direction, which is desirable for accurate dense vector prediction. The dense vectors are generated by an end-to-end trainable encoder $\textbf{Encoder}$, which is optimized during the training process. This ensures that the dense vectors are dynamically refined and adapted to the specific requirements of the recommendation task.

The overall loss function is formulated as:
\begin{equation}
\mathcal{L} = \mathcal{L}_{\text{sparse}} + \mathcal{L}_{\text{dense}}
\end{equation}
The dual-objective loss function enables a balanced optimization process, where the model dynamically refines dense vectors guided by sparse IDs. This end-to-end training approach captures both high-level semantics and feature-level information, optimizing sparse and dense representations jointly for superior performance.
\subsection{Coarse-to-Fine Generation}

\begin{figure}[htbp]
\centering
\includegraphics[width=0.45\textwidth]{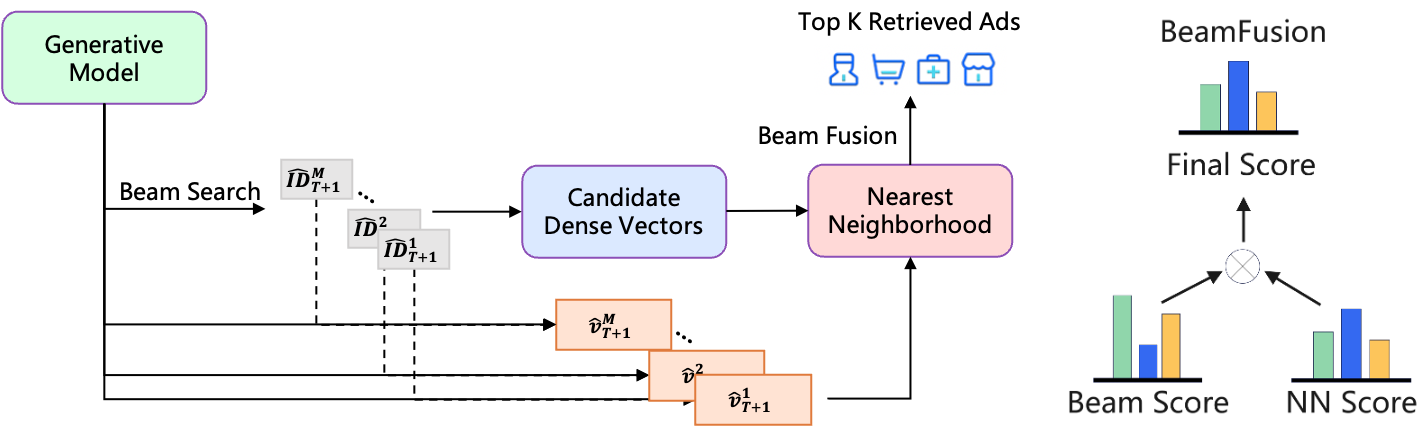}
\caption{Illustration of the Coarse-to-Fine Generation process. During inference, $M$ sparse IDs are generated via Beam Search, and appended to the sequence. Dense vectors are then generated and used in ANN to obtain candidate items. BeamFusion combines beam scores and similarity scores to rank candidates, from which the top $K$ items are selected.}
\label{fig:inference_with_beamfusion}
\end{figure}
During the inference phase, COBRA implements the coarse-to-fine generation procedure, involving the sequential generation of sparse IDs followed by the refinement of dense vectors in a cascaded manner, as illustrated in Figure~\ref{fig:inference_with_beamfusion}. The coarse-to-fine generation process in COBRA is designed to capture both the categorical essence and fine-grained details of user-item interactions. This process involves two main stages:

\textbf{Sparse ID Generation}: Given a user sequence $S_{1:T}$, we utilize the ID probability distribution modeled by the Transformer Decoder, $\hat{ID}_{T+1} \sim P(i_{T+1} | S_{1:T})$, and employ the BeamSearch algorithm to derive the top $M$ IDs. The formulation is as follows:

\begin{equation}
\{\hat{\mathbf{ID}}_{T+1}^k\}_{k=1}^{M} = \text{BeamSearch}(\textbf{TransformerDecoder}(\mathbf{S}_{1:T}), M)
\end{equation}
where $k \in \{1, 2, \dots, M\}$. Each generated ID is associated with a beam score $\phi_{\hat{\mathbf{ID}}_{T+1}^k}$.

\textbf{Dense Vector Refinement}: Each generated sparse ID $ \hat{\mathbf{ID}}_{T+1}^k $ is subsequently converted into an embedding and appended to the previous cascaded sequence embedding $\mathbf{S}_{1:T}$. Then the corresponding dense vector $ \hat{\mathbf{v}}_{T+1}^k $ is generated:
\begin{equation}
\hat{\mathbf{v}}_{T+1}^k = \textbf{TransformerDecoder}([\mathbf{S}_{1:T}, \textbf{Embed}(\hat{\mathbf{ID}}_{T+1}^k)])
\end{equation}

After that, we employ Approximate Nearest Neighbor (ANN) search to retrieve the top $ N $ candidate items:
\begin{equation}
\mathcal{A}_k = \text{ANN}(\hat{\mathbf{ID}}_{T+1}^k, \mathcal{C}(\hat{\mathbf{ID}}_{T+1}^k), N)
\end{equation}
where $ \mathcal{C}(\hat{\mathbf{ID}}_{T+1}^k) $ is the set of candidate items associated with sparse ID $ \hat{\mathbf{ID}}_{T+1}^k $, and $ N $ represents the number of top items to be retrieved.

\textbf{BeamFusion Mechanism}
In order to achieve a balance between precision and diversity, we devise a globally comparable score for items corresponding to each sparse ID. This score is capable of reflecting both the differences among different sparse IDs and the fine-grained difference among items under the same sparse ID. To accomplish this, we propose the BeamFusion mechanism:
\begin{equation}
\Phi^{(\hat{\mathbf{v}}_{T+1}^k, \hat{\mathbf{ID}}_{T+1}^k, \mathbf{a})} = \text{Softmax}(\tau \phi_{\hat{\mathbf{ID}}_{T+1}^k}) \times \text{Softmax}(\psi \cos(\hat{\mathbf{v}}_{T+1}^k, \mathbf{a}))
\end{equation}
where $\mathbf{a}$ represents the candidate item, $\tau$ and $\psi$ are coefficients, and $\phi_{\hat{\mathbf{ID}}_{T+1}^k}$ denotes the beam score obtained during the beam search process.

Finally, we rank all candidate items based on their BeamFusion Scores and select the top $ K $ items as the final recommendations:
\begin{equation}
\mathcal{R} = \text{TopK}\left(\bigcup_{k=1}^{M} \mathcal{A}_k, \Phi, K\right)
\end{equation}
where $ \mathcal{R} $ denotes the set of final recommendations, and $ \text{TopK} $ represents the operation of selecting the top $ K $ items with the highest BeamFusion Scores.
\section{Experiment}
\label{sec:experiment}
This section presents a comprehensive evaluation of the COBRA framework using both public and industrial datasets. Our experiments focus on assessing COBRA's ability to improve recommendation accuracy and diversity, while also validating its practical effectiveness through offline and online evaluations.

\begin{table}[h]
    \centering
    \begin{tabular}{|l|cc|c|c|}
        \hline
        Dataset & \# Users & \# Items & \multicolumn{2}{c|}{Sequence Length} \\
        \hline
        & & & Mean & Median \\
        \cline{4-5}
        Beauty & 22,363 & 12,101 & 8.87 & 6 \\
        Sports and Outdoors & 35,598 & 18,357 & 8.32 & 6 \\
        Toys and Games & 19,412 & 11,924 & 8.63 & 6 \\
        \hline
    \end{tabular}
    \caption{Dataset Statistics}
    \label{tab:dataset_stats}
\end{table}

\begin{table}[htbp]
\centering
\caption{Performance comparison on public datasets. The best metric for each dataset is highlighted in bold, while the second-best is underlined.}
\label{tab:public_performance}
\begin{tabularx}{0.5\textwidth}{@{}l|p{0.12\textwidth}|>{\centering\arraybackslash}X>{\centering\arraybackslash}X>{\centering\arraybackslash}X>{\centering\arraybackslash}X@{}}
\toprule
 & \textbf{Method} & \textbf{R@5} & \textbf{N@5} & \textbf{R@10} & \textbf{N@10} \\
\midrule
\multirow{9}{*}{\rotatebox[origin=c]{90}{Beauty}} & P5 & 0.0163 & 0.0107 & 0.0254 & 0.0136 \\
 & Caser & 0.0205 & 0.0131 & 0.0347 & 0.0176 \\
 & HGN & 0.0325 & 0.0206 & 0.0512 & 0.0266 \\
 & GRU4Rec & 0.0164 & 0.0099 & 0.0283 & 0.0137 \\
 & BERT4Rec & 0.0203 & 0.0124 & 0.0347 & 0.0170 \\
 & FDSA & 0.0267 & 0.0163 & 0.0407 & 0.0208 \\
 & SASRec & 0.0387 & 0.0249 & 0.0605 & 0.0318 \\
 & $\text{S}^3$-Rec & 0.0387 & 0.0244 & 0.0647 & 0.0327 \\
 & TIGER & \underline{0.0454} & \underline{0.0321} & \underline{0.0648} & \underline{0.0384} \\
 & \textbf{COBRA[Ours]} & \textbf{0.0537} & \textbf{0.0395} & \textbf{0.0725} & \textbf{0.0456} \\
\midrule
\multirow{9}{*}{\rotatebox[origin=c]{90}{Sports}} & P5 & 0.0061 & 0.0041 & 0.0095 & 0.0052 \\
 & Caser & 0.0116 & 0.0072 & 0.0194 & 0.0097 \\
 & HGN & 0.0189 & 0.0120 & 0.0313 & 0.0159 \\
 & GRU4Rec & 0.0129 & 0.0086 & 0.0204 & 0.0110 \\
 & BERT4Rec & 0.0115 & 0.0075 & 0.0191 & 0.0099 \\
 & FDSA & 0.0182 & 0.0122 & 0.0288 & 0.0156 \\
 & SASRec & 0.0233 & 0.0154 & 0.0350 & 0.0192 \\
 & $\text{S}^3$-Rec & 0.0251 & 0.0161 & 0.0385 & 0.0204 \\
 & TIGER & \underline{0.0264} & \underline{0.0181} & \underline{0.0400} & \underline{0.0225} \\
 & \textbf{COBRA[Ours]} & \textbf{0.0305} & \textbf{0.0215} & \textbf{0.0434} & \textbf{0.0257} \\
\midrule
\multirow{9}{*}{\rotatebox[origin=c]{90}{Toys}} & P5 & 0.0070 & 0.0050 & 0.0121 & 0.0066 \\
 & Caser & 0.0166 & 0.0107 & 0.0270 & 0.0141 \\
 & HGN & 0.0321 & 0.0221 & 0.0497 & 0.0277 \\
 & GRU4Rec & 0.0097 & 0.0059 & 0.0176 & 0.0084 \\
 & BERT4Rec & 0.0116 & 0.0071 & 0.0203 & 0.0099 \\
 & FDSA & 0.0228 & 0.0140 & 0.0381 & 0.0189 \\
 & SASRec & 0.0463 & 0.0306 & 0.0675 & 0.0374 \\
 & $\text{S}^3$-Rec & 0.0443 & 0.0294 & 0.0700 & 0.0376 \\
 & TIGER & \underline{0.0521} & \underline{0.0371} & \underline{0.0712} & \underline{0.0432} \\
 & \textbf{COBRA[Ours]} & \textbf{0.0619} & \textbf{0.0462} & \textbf{0.0781} & \textbf{0.0515} \\
\bottomrule
\end{tabularx}
\end{table}

\begin{table*}[htbp]
\centering
\caption{Performance comparison on industrial dataset}
\label{tab:industrial_performance}
\begin{tabularx}{0.6\textwidth}{@{}p{0.2\textwidth}|ccccc@{}}
\toprule
\textbf{Method} & \textbf{R@50} & \textbf{R@100} & \textbf{R@200} & \textbf{R@500} & \textbf{R@800} \\
\midrule
\textbf{COBRA} & \textbf{0.1180} & \textbf{0.1737} & \textbf{0.2470} & \textbf{0.3716} & \textbf{0.4466} \\
COBRA w/o ID & 0.0611 & 0.0964 & 0.1474 & 0.2466 & 0.3111 \\
COBRA w/o Dense & 0.0690 & 0.1032 & 0.1738 & 0.2709 & 0.3273 \\
COBRA w/o BeamFusion & 0.0856 & 0.1254 & 0.1732 & 0.2455 & 0.2855 \\
\bottomrule
\end{tabularx}
\end{table*}

\subsection{Public Dataset Experiments}

\subsubsection{Datasets and Evaluation Metrics}
In our experiments, we evaluate the performance of COBRA using the Amazon Product Reviews dataset~\cite{amazon,amazon_2}, which is a well-established benchmark for recommendation tasks. This dataset encompasses product reviews and associated metadata collected from May 1996 to September 2014. Our analysis focuses on three specific subsets: "Beauty," "Sports and Outdoors," and "Toys and Games." For each subset, we construct item embeddings leveraging attributes such as title, price, category, and description. To ensure data quality, we apply a 5-core filtering process, eliminating items with fewer than five user interactions and users with fewer than five item interactions. Detailed statistics of the datasets are presented in Table~\ref{tab:dataset_stats}.
For the evaluation of recommendation accuracy and ranking quality, we employ Recall@K and NDCG@K, specifically at $K = 5$ and $K = 10$. These metrics provide insights into the system's ability to accurately recommend relevant items and maintain a high-quality ranking order.

\begin{figure*}[htbp]
\centering
\begin{subfigure}[t]{0.32\textwidth}
    \centering
    \includegraphics[width=\textwidth]{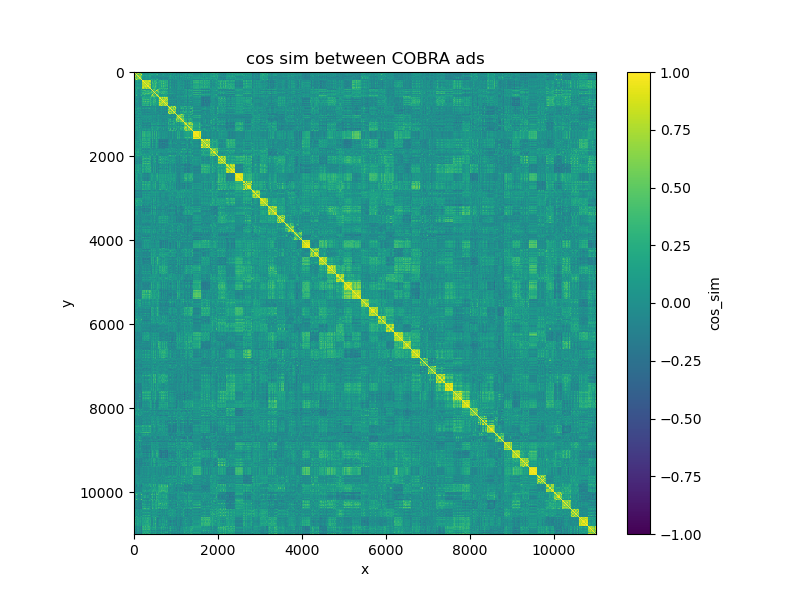}
    \caption{COBRA}
    \label{fig:similarity_matrix_cobra}
\end{subfigure}
\hfill
\begin{subfigure}[t]{0.32\textwidth}
    \centering
    \includegraphics[width=\textwidth]{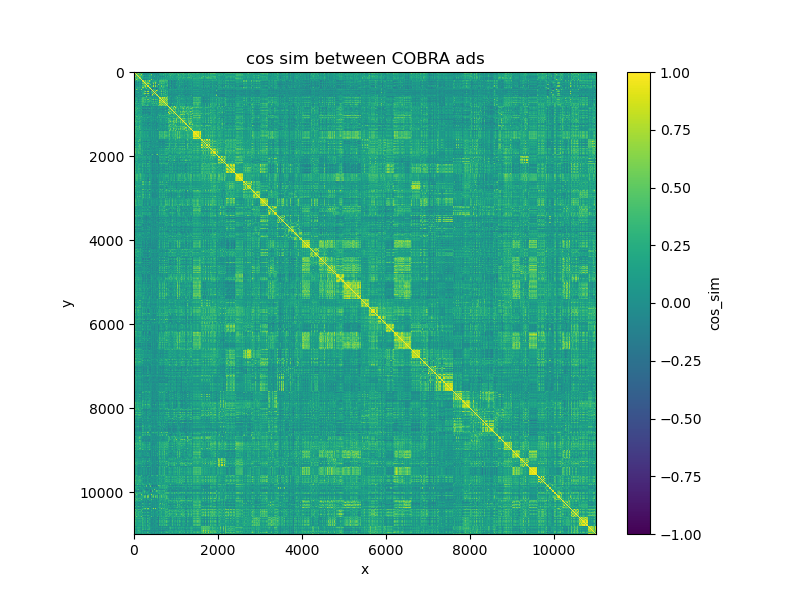}
    \caption{COBRA w/o ID}
    \label{fig:similarity_matrix_cobra_woid}
\end{subfigure}
\hfill
\begin{subfigure}[t]{0.32\textwidth}
    \centering
    \includegraphics[width=\textwidth]{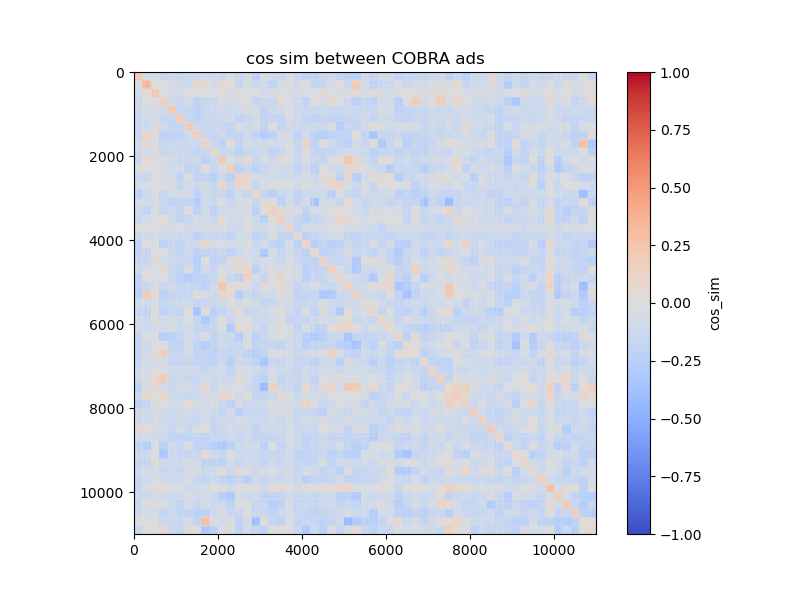}
    \caption{Difference}
    \label{fig:similarity_matrix_difference}    
\end{subfigure}
\caption{Cosine similarity matrices for advertisement dense embeddings. (a) COBRA's dense embeddings exhibit strong intra-ID cohesion and inter-ID separation. (b) COBRA w/o ID shows weaker category separation. (c) The difference matrix quantifies the enhancement in cohesion and separation when sparse IDs are incorporated.}
\label{fig:similarity_matrix}
\end{figure*}

\subsubsection{Baselines}
To comprehensively evaluate the performance of our proposed COBRA method, we compare it with several state-of-the-art recommendation methods:
\begin{itemize}
    \item \textbf{P5}~\cite{p5}: Transforms recommendations into natural language sequences.
    \item \textbf{Caser}~\cite{caser}: Captures sequential patterns using convolutional layers.
    \item \textbf{HGN}~\cite{hgn}: Hierarchical gating networks for long-/short-term user interests.
    \item \textbf{GRU4Rec}~\cite{gru4rec}: Models user behavior with gated recurrent units.
    \item \textbf{SASRec}~\cite{Sasrec}: Transformer-based model for long-term dependencies.
    \item \textbf{FDSA}~\cite{fdsa}: Self-attentive model for item-feature transitions.
    \item \textbf{BERT4Rec}~\cite{bert4rec}: Bidirectional self-attention with cloze objective.
    \item \textbf{$\text{S}^3$-Rec}~\cite{s3rec}: Contrastive learning for recommendation.
    \item \textbf{TIGER}\cite{tiger}: Uses RQ-VAE to encode item content features and Transformer for generative retrieval.
\end{itemize}
These methods are chosen to represent a variety of recommendation techniques, including sequential dense recommendation and generative recommendation.

\subsubsection{Implementation Details}
In our approach, we adopt a method for generating semantic IDs similar to the one used in \cite{tiger}. However, unlike \cite{tiger}, which uses a different configuration, we employ a 3-level semantic ID structure, where each level corresponds to a codebook size of 32. These semantic IDs are generated using the T5 model. COBRA is implemented with a lightweight architecture, featuring a 1-layer encoder and a 2-layer decoder.

\subsubsection{Results}

COBRA consistently surpasses all baseline models across various metrics, as presented in Table~\ref{tab:public_performance}. On the "Beauty" dataset, COBRA achieves a Recall@5 of 0.0537 and a Recall@10 of 0.0725, exceeding the previous best model (TIGER) by 18.3\% and 11.9\%, respectively. For the "Sports and Outdoors" dataset, COBRA records a Recall@5 of 0.0305 and an NDCG@10 of 0.0215, outperforming TIGER by 15.5\% and 18.8\%, respectively. On the "Toys and Games" dataset, COBRA attains a Recall@10 of 0.0462 and an NDCG@10 of 0.0515, surpassing TIGER by 24.5\% and 19.2\%, respectively.

\begin{figure}[htbp]
\centering
\includegraphics[width=.49\textwidth]{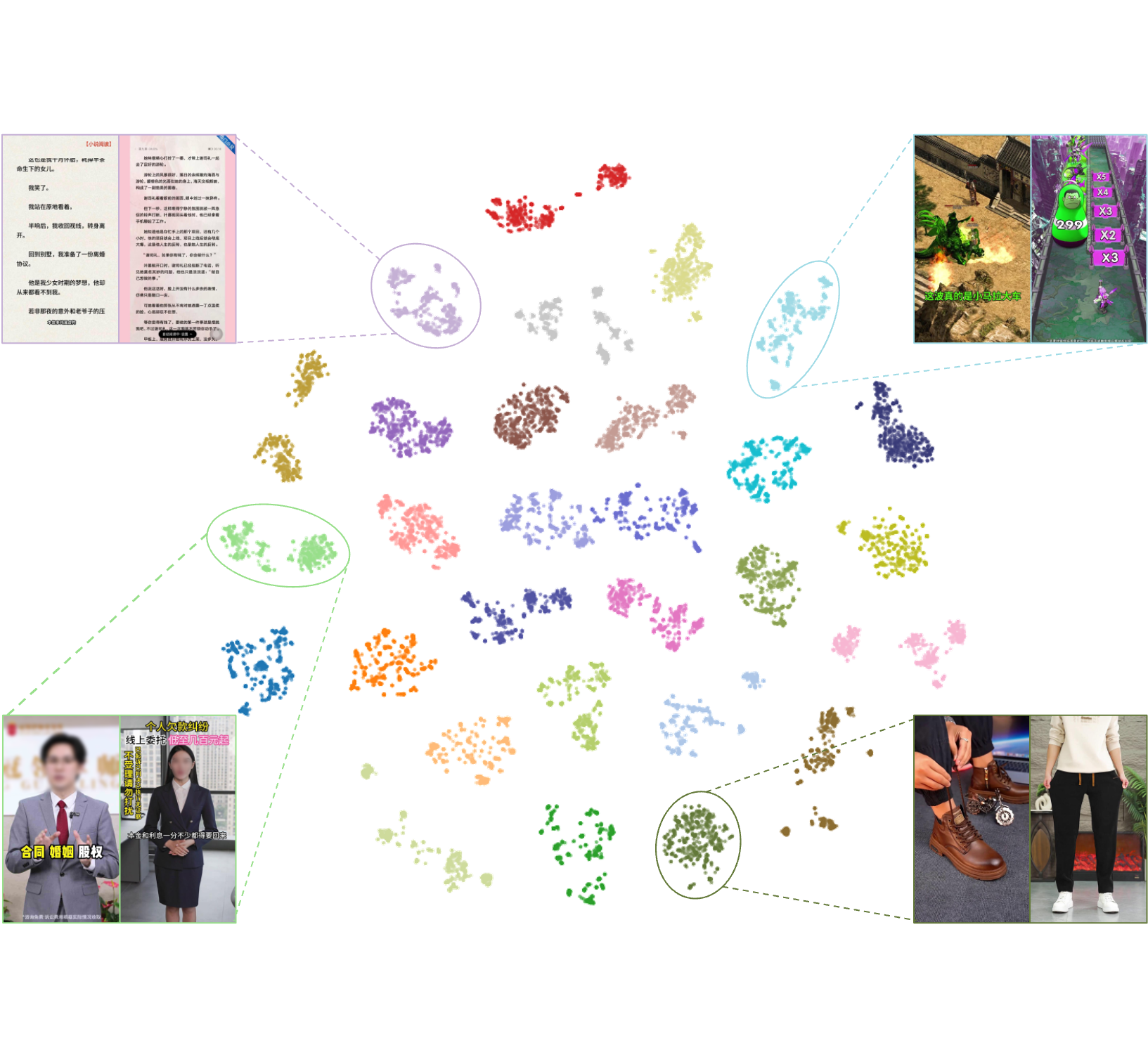}
\caption{Embedding Visualization using t-SNE. The plot illustrates the distribution of 10,000 randomly sampled advertisement embeddings in a two-dimensional space for COBRA. Distinct clustering centers are observed for various IDs.}
\label{fig:embedding_visualization}
\end{figure}

\begin{figure}[htbp]
\centering
\includegraphics[width=.49\textwidth]{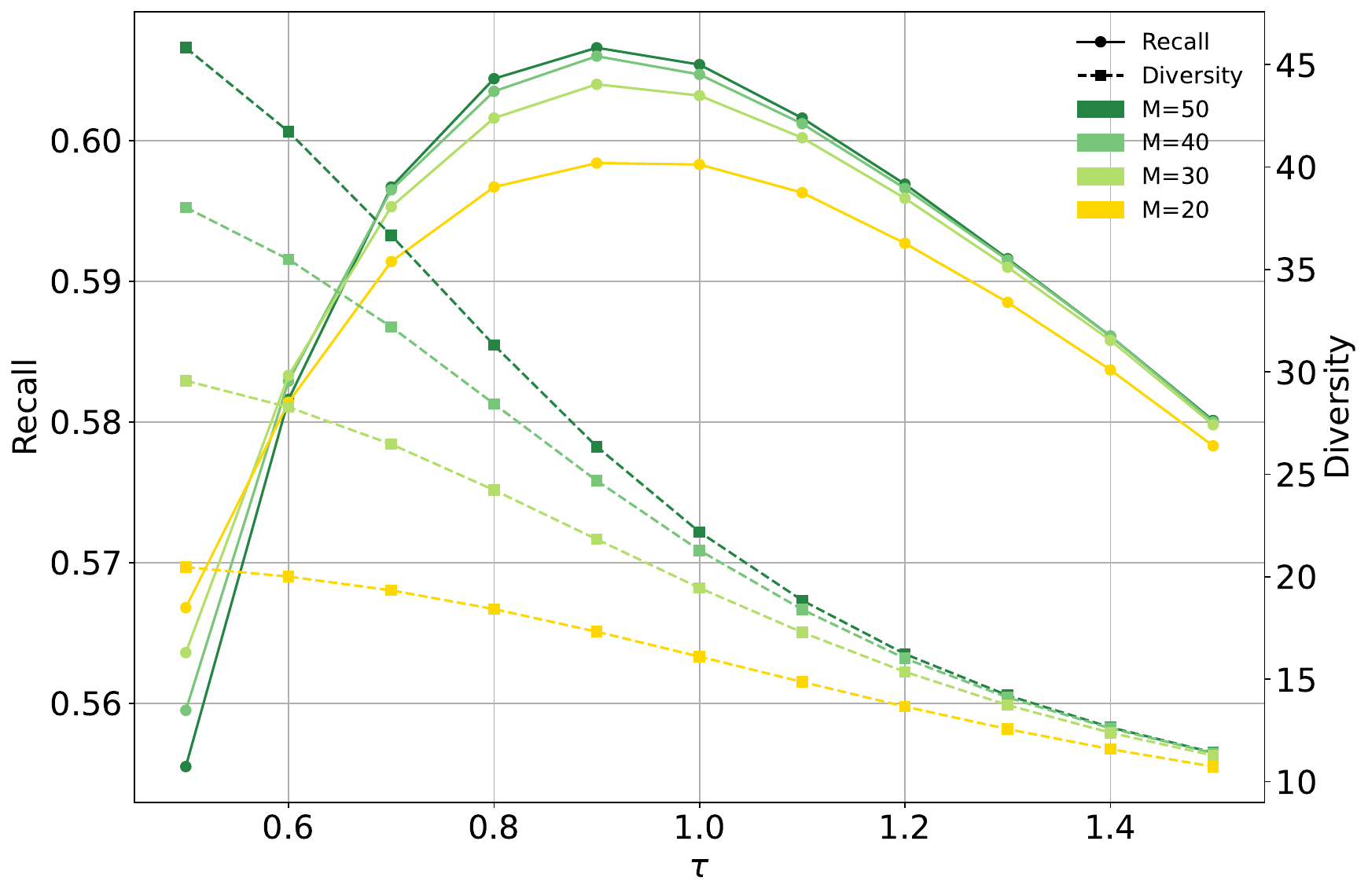}
\caption{\textbf{Recall-Diversity Curves.} The x-axis represents the coefficient $\tau$, and the y-axis shows the Recall@2000 and Diversity metrics.}
\label{fig:recall_diversity}
\end{figure}

\subsection{Industrial-scale Experiments}

\subsubsection{Dataset and Evaluation Metrics}

To comprehensively evaluate the proposed COBRA method, we conduct experiments on the Baidu Industrial Dataset, a large-scale dataset derived from user interaction logs on the Baidu advertising platform. The dataset encompasses diverse recommendation scenarios, including list-page, dual-column, and short-video. It consists of five million users and two million advertisements, providing a comprehensive representation of real-world user behavior and advertising content.
Advertisers and advertisements are represented through attributes such as title, industry labels, brand, and campaign text. These attributes are processed and encoded into two-level sparse IDs and dense vectors, capturing both coarse-grained and fine-grained semantic information. This dual representation enables COBRA to effectively model user preferences and item characteristics.
The dataset is divided into two parts: $D_{\text{train}}$ and $D_{\text{test}}$. The training set, $D_{\text{train}}$, includes user interaction logs collected over the first 60 days, covering recommendation content interactions during this period. The test set, $D_{\text{test}}$, is constructed from logs of the subsequent day following the $D_{\text{train}}$ period, serving as a benchmark to assess model performance.
For offline evaluation, we employ Recall@K as the evaluation metric, testing with $K \in \{50, 100, 200, 500, 800\}$. This metric provides a measure of the model's ability to accurately retrieve relevant recommendations at various thresholds.

\subsubsection{Baselines}
We compare COBRA against its variants:
\begin{itemize}
    \item \textbf{COBRA w/o ID}: Removes sparse IDs, relying solely on dense vectors. This variant resembles RecFormer~\cite{recformer}, using lightweight transformers for sequence modeling.
    \item \textbf{COBRA w/o Dense}: Removes dense vectors, using only sparse IDs. Due to the coarse-grained nature of IDs, this variant adopts a generative retrieval method similar to TIGER~\cite{tiger}, leveraging semantic IDs for retrieval.
    \item \textbf{COBRA w/o BeamFusion}: Removes the BeamFusion module, using top-1 sparse ID and nearest-neighbor retrieval for top-$k$ results.
\end{itemize}

\subsubsection{Implementation Details}

COBRA is built upon a Transformer-based architecture. In this framework, the text encoder processes advertisement text into sequences, which are then handled by the sparse ID head to predict 2-level semantic IDs configured as $32 \times 32$. For more fine-grained modeling of advertisements, the variant COBRA w/o Dense employs 3-level semantic IDs ($256 \times 256 \times 256$).

\subsubsection{Results}

As shown in Table~\ref{tab:industrial_performance}, COBRA consistently outperforms all its variants across all evaluated metrics. At $K = 500$, COBRA achieves a Recall@$500$ of 0.3716, representing a 42.2\% improvement over the COBRA w/o Dense variant. When $K = 800$, COBRA attains a Recall@$800$ of 0.4466, reflecting a 43.6\% improvement over the COBRA w/o ID variant and a 36.1\% enhancement compared to COBRA w/o BeamFusion. Notably, at relatively smaller values of $K$, the absence of Dense or ID representations results in more pronounced performance declines, underscoring the importance of cascaded representations for achieving granularity and precision. Conversely, as the recall size $K$ increases, the performance advantages associated with BeamFusion become increasingly evident, demonstrating its effectiveness in practical industrial recall systems.

The results further underscore the contributions of specific components:

\begin{itemize}
    \item Excluding sparse IDs (COBRA w/o ID) leads to a recall reduction ranging from 26.7\% to 41.5\%, highlighting the critical role of semantic categorization.
    \item The removal of 3-level semantic IDs (COBRA w/o Dense) results in a performance drop between 30.3\% and 48.3\%, underscoring the importance of fine-grained modeling.
    \item Eliminating BeamFusion results in a recall decrease of 27.5\% to 36.1\%, emphasizing its significance in the integration of sparse signals.
\end{itemize}

\subsection{Further Analysis}

\subsubsection{Analysis of Representation Learning}
To evaluate the representation learning capabilities of the COBRA model, we construct similarity matrices for the dense embeddings of advertisements, as illustrated in Figure~\ref{fig:similarity_matrix}. The COBRA model exhibits significant intra-ID cohesion and inter-ID separation, as demonstrated in the top heatmap of Figure~\ref{fig:similarity_matrix_cobra}. This suggests that COBRA's dense embeddings proficiently capture detailed item characteristics while preserving semantic consistency within categories. Conversely, the model variant without sparse IDs (Figure~\ref{fig:similarity_matrix_cobra_woid}) shows weaker category separation, underscoring the importance of sparse IDs in maintaining semantic structure. The difference matrix in Figure~\ref{fig:similarity_matrix_difference} quantitatively confirms that incorporating sparse IDs enhances both cohesion and separation.

Further validation of COBRA's embeddings is achieved through visualizing the distribution of advertisement embeddings in a two-dimensional space using t-SNE. By randomly sampling 10,000 advertisements, distinct clustering centers for various categories are observed. Figure~\ref{fig:embedding_visualization} reveals that advertisements are effectively clustered by category, indicating strong cohesion within categories. The clusters in purple, teal, light green, and dark green correspond primarily to advertisements for novels, games, legal services, and clothing, respectively. This demonstrates that the advertisement representations effectively capture semantic information.

\subsubsection{Recall-Diversity Equilibrium}
Balancing accuracy and diversity is a challenge in the retrieval stage of recommendation systems. To analyze this trade-off in COBRA, we examine recall-diversity curves, which depict how recall and diversity metrics evolve with the coefficient $\tau$ in the BeamFusion mechanism under a fixed $\phi = 16$. Specifically, the curves in Figure~\ref{fig:recall_diversity} illustrate how Recall@2000 and diversity change as $\tau$ varies.As shown in the figure, increasing \( \tau \) generally leads to a decrease in diversity. COBRA achieves an optimal balance between recall and diversity at $\tau = 0.9$ and $\phi = 16$. Here, the model maintains high accuracy while ensuring that the recommendations cover a sufficiently diverse set of items. The diversity metric, defined as the number of different IDs in the recalled items, reflects the model's ability to avoid redundancy and provide users with a broader range of options. 
This fine-grained control over $\tau$ and $\phi$ allows practitioners to adjust the emphasis on accuracy or diversity based on specific business objectives. For instance, platforms prioritizing exploration can decrease $\tau$ to enhance diversity. This flexibility distinguishes COBRA from models with fixed retrieval strategies, making it adaptable to diverse recommendation scenarios.

\subsection{Online Results}
To validate COBRA's real-world effectiveness, we conducted online A/B tests on the Baidu Industrial Dataset in January 2025. The test covered 10\% of user traffic, ensuring statistical significance. The primary evaluation metrics were conversion and Average Revenue Per User (ARPU), which directly reflect user engagement and economic value.
In the field covered by our proposed strategy, COBRA achieved a 3.60\% increase in conversion and a 4.15\% increase in ARPU. The results demonstrate that COBRA's hybrid architecture not only enhances recommendation quality in offline evaluations but also drives measurable business outcomes in production environments.
\section{Conclusions}
In this work, we introduced COBRA, a generative recommendation framework that integrates cascaded sparse and dense representations for improved accuracy and diversity. COBRA employs a coarse-to-fine generation process, first generating sparse ID to capture the categorical essence of an item, then refining it with a dense vector. 
Extensive experiments demonstrate that COBRA outperforms state-of-the-art methods in both accuracy and diversity. Evaluations on public and industrial datasets, along with online A/B tests, confirm its effectiveness and practical applicability. By leveraging both sparse and dense representations, COBRA offers a robust solution for large-scale recommendation tasks.




\end{document}